\documentclass[twocolumn,showpacs,amssymb,floatfix,aps]{revtex4}

\usepackage{graphicx}% Include figure files
\usepackage{dcolumn}% Align table columns on decimal point
\usepackage{bm}% bold math
\usepackage{epsfig}

\begin{document}
\title{Interference of Electromagnetic waves in the background of the gravitational waves }
\author{Wenyu Wang, Yang Xu}
\affiliation{Institute of Theoretical Physics, College of Applied Science,
Beijing University of Technology, Beijing 100124, China}

\date{\today}
\begin{abstract}
Based on the relationship between proper distance and coordinate distance,
the geometrical phenomenon caused by the passing gravitational waves
can not be observed  locally. The electromagnetic wave equations
in the background gravitational waves are studied. We find that the expansion and contraction
of wave lengths are always synchronous with the objects it measures.
The background of the gravitational waves leads to dissipation and dispersion in the propagation of
electromagnetic wave. The phase of the gravitational waves control the
dissipation term  and dispersion  term in the telegrapher's equation.
The linearly polarized laser beam propagating in the direction of the incoming gravitational waves
can give a possible measurement on the local metric.  In case of the pulsed beats passing by,
the relaxation time is greater than the period of the gravitational waves,
thus the detector may only show a signal of the modulation of the beats.
Finally we proposed  a non-local interference experiment
to detect the high-frequency gravitational waves. It is similar to the measurement
of redshift caused by gravitation. Together with the ordinary detector,
it will give us further and mutually measurements of the gravitational waves.
\end{abstract}
\maketitle

\section{Introduction}
The detection of gravitational waves (GWs)  by the Laser Interferometer
Gravitational-wave Observatory (LIGO) opened a new era of gravitational-wave astronomy
\cite{TheLIGOScientific:2016agk}. The precise wave shape of the first
GWs signal is consistent with the waves shape produced from the coalescence
of two black holes  predicted by the numerical simulation
in the theory of the general relativity \cite{Vitale:2016rfr}.
Later, several other GWs events were observed and studied in detail by the community.
Most of these events were identified as the merging of binary black hole or neutron star
\cite{Abbott:2016nmj, Abbott:2017vtc, Abbott:2017oio, Abbott:2017gyy}, i.e.
the detection of GW170814 provides an interesting approach of probing
the polarization modes of the GWs \cite{Abbott:2017oio},
GW170817 and GRB 170817A showed a very tight bound on
the speed of the GWs \cite{TheLIGOScientific:2017qsa, Cabral:2016klm}.
As a new window to explore our universe, the GWs play an important role in modern cosmology,
such as black hole physics \cite{Gondan:2018khr}, tests of strong gravitation field regime,
other extended theory of gravitation and  the early universe \cite{Goldstein:2017mmi}.
Based on ingenious experiments and advanced technologies,
an extraordinary amount of analysis has been carried out
towards higher precision tests of relevant issues.

Besides LIGO experiment, there are several different types of the
GWs detectors aiming at different frequency band of GWs. Pulsar timing arrays (PTAs)
\cite{Kramer:2013kea, McLaughlin:2013ira, Hobbs:2013aka} detect GWs in the lower-frequency band.
Meanwhile, Advanced LIGO \cite{Harry:2010zz},
Advanced Virgo \cite{TheVirgo:2014hva} and KAGRA \cite{Aso:2013eba},
these ground-based interferometers detect the GWs in the high-frequency band ($10\sim10^4$ Hz),
possessing higher sensitivity over a broader frequency band.
The theoretical principle of detection method is a simple application
of general relativity theory, the GWs which is far from the source can be
treated as a time-dependent perturbation of the space-time metric,
consequently causing length changing with respect to its polarizations.
For example, Advanced LIGO measures linearly differential displacement along
the arms which is proportional to the amplitude of the GWs.
By means of the Michelson interferometer, the ground-based detectors
have a much higher accuracy on detecting the small displacement.
Any length changing in two arms of the interferometer will produce
a phase shift, thus cause changes of interference pattern  at
the output of the detector \cite{TheLIGOScientific:2014jea}.
Therefore, the Advanced LIGO detectors translate strain into a measurable
optical signal \cite{signal}.

In flat space-time, interferometers are widely used in science and
industry for the measurement of small displacements.
However, the principle for measuring proper distance in the background GWs
is quite questionable. According to the
general relativity, the local physical measurement of space-time
can not detect the influence of the background metric,
only global comparison can show the difference between
areas of space-time with different local metric.
This is because that the gravitational field affects space and time
standards in exactly the same way as it affects the object being studied \cite{SW}.
This properties of space-time metric have been studied
in many papers and textbooks, but the application of interferometers
in curved space-time is hardly mentioned.  Thus to confirm the existence of the GWs,
the mechanism that using the laser interference needs to be subtly checked.
In the interference measurement, the laser is assumed as a flux of photon
with a speed of light which is unaffected by the GWs. The measurement
of  the intensity variation can give the  phase shift of the electromagnetic waves,
then the displacements of the length of the arms in the interferometer.
However, it is controversial for using "laser ruler"
to measure length when it has been assumed that the ruler (wave length of the
laser)is unaffected. The propagation of photon are in fact obey the classical
electromagnetic theory. If we want to make sure about the phase shift,
the solver of electromagnetic wave in the curved space time is needed.
Only the electromagnetic waves equation under the background GWs
can give a decisive judgement  in revealing the characteristic of laser light.

In all, we will explore the electromagnetic wave equation and its interference
in the interferometers in the curved space time with background GWs in this paper,
The implications and comparisons with the current detection results of the GWs are also
discussed. The paper is organized as following. In sec. \ref{sec2},
the coordinate distance and proper distance, and some basic concepts of the GWs are illustrated.
we also discussed the method and the common understanding
of the phase shift measured by the LIGO experiment.
In sec. \ref{sec3}, we study the electromagnetic wave equation and the solution
in a background gravitational wave, the interference of the electromagnetic waves
and its relation with the GWs are also discussed. In sec. \ref{sec4}, we show our result
of the phase shift and compare it with the LIGO's. We also try to show a new
method of measurement of the GWs. The conclusion is in given the sec. \ref{sec5}.

\section{space-time metric and the gravitational waves}\label{sec2}
In the framework of the general relativity, or any other metric
based gravitational theories\cite{extended},
equivalence principle implies that the phenomenon of gravitation can be considered
as a specific nonlinearly coordinate transformations of space-time.
Every physical theory can be generalized to a theory in the
curved space  by the general covariance principle  which
implies that the theory should be  coordinate independent.
According to general relativity, material fields will affect the measure
of space-time via Einstein's equation. In four dimensional space-time,
the metric tensor $g_{\mu\nu}$ is defined to measure
space-time interval square $ds^2$ in a given frame  ${x^\mu}$,
\begin{eqnarray}
ds^2=g_{\mu\nu}dx^\mu dx^\nu,\label{metric}
\end{eqnarray}
with symmetric indices $\mu\nu$. $\mu$ and $\nu$ equal to  $0,~1,~2,~3$
which correspond to the coordinates $t,~x,~y,~z$ as the usual convention
in the literature. Generally, the space-time metric has 10 components,
which define the measure of space-time. It is the basic dynamical variable in
the  gravitational theories, such as general relativity. There are many experimental
tests of the variation of metric induced by the gravitational field, such as
gravitational redshift, precession of planetary perihelion,
gravitational deflection of light and time delay.
The GWs is an important prediction of the theory general relativity.
When the GWs pass through the earth, the metric will slightly deviate from flat space-time.
The detection of the GWs also can be viewed as the direct measurements on
the space components of metric\cite{GAS}.

\begin{figure}
\scalebox{0.3}{\epsfig{file=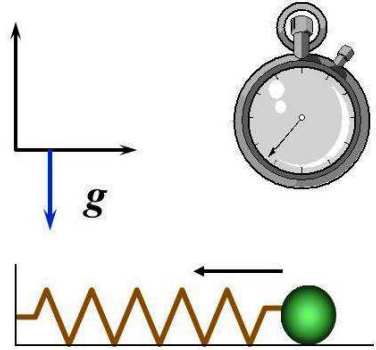}}
\caption{The harmonic oscillator and a relatively static clock are always synchronous,
no matter whether they are rest in the gravitational field or falling freely in it.}
\label{fig1}
\end{figure}
We find that understanding the relation between proper interval and  coordinate interval
is very important in the issue we are talking about in this paper.
From equation (\ref{metric}), we can see that the proper distance differential
interval $dX$ along X direction is
\begin{eqnarray}
dX&=&\sqrt{-g_{11}}dx,\label{lxdx}
\end{eqnarray}
in which $dx$ is the coordinate differential. In relativity theory, one should keep in mind that
the proper interval $dX$ and proper time interval are independent of frame of reference, they
are actually the observable corresponding to a real physical process in the local place,
such as a simple harmonic motion, stimulated radiation of atoms, {\em et. al.}
While the coordinate interval $dx$ is globally established but it is in fact unmeasurable.
Any local measurements will always indicates its proper distance
for the measurements taken by the local ruler or clock.
This circumstance is demonstrated by a sketch graph as shown in
the FIG. \ref{fig1} in which a clock is relative rest to harmonic oscillator.
We can assume that the period of
oscillator equals to the interval of the clock which may be recorded by
other physical periodic processes. As talked in the introduction, a local measurement
can not show the variation of the metric, the gravitational field affects space and time
standards in exactly the same way as it affects the oscillator being studied and the internal
periodic process happening in the clock. Whether they are rest in the gravitational field or
falling freely in it. The proper time of the clock will always shows the
period of the harmonic oscillator with exactly the same value,
as long as the observation is disposed locally.
This means that there is no effect of time dilation, change of length can be observed at this
local place. To make this effect detectable, one must compare the clocks placed
in different location. This can also be understood by virtue of the equivalence principle,
the free falling observer will never experience a time dilation,
such phenomenon can be detected by a relatively static observer.
Note that, we often treat coordinate interval $dx$ as being observed far away from
the strong gravitational source, where the space-time is Minkowskian.

This understanding of the proper interval and the coordinate
 is crucial for LIGO detectors, since the experiment
is taken place by using a local laser. In this case, the wavelength is in fact the local ruler,
because the wavelength of laser determines the interference patten in the interferometer.
As talked in the above paragraph, it can not show the changing  of the metric
 without a global comparison. However, things are different, this laser ruler
is an electromagnetic wave, thus the details of wave equations which are influenced by GWs
 should be checked. And what we should ask is  whether
the wavelength change under the GWs? And what is detected by LIGO experiment exactly?

Before checking the wavelength of the laser, let's take a short review on the GWs in the framework of
the general relativity. The metric of the weak GWs $h_{\mu\nu}$ (a small quantity)
in vacuum is the solution of Einstein equation
\begin{eqnarray}
\partial^2\phi_{\mu\nu}&=&0\label{wave eq},
\end{eqnarray}
where \cite{SW}
\begin{eqnarray}
\phi_{\mu\nu}&=&h_{\mu\nu}-\frac{1}{2}\eta_{\mu\nu}h.
\end{eqnarray}
$\eta_{\mu\nu}=$diag$(1,-1,-1,-1)$ is Minkowskian metric,
\begin{eqnarray}
h=h_{\mu\nu}\eta^{\mu\nu}
\end{eqnarray}
is the trace of the GWs metric. The gauge condition of $\phi_{\mu\nu}$ is
\begin{eqnarray}
\partial^\mu\phi_{\mu\nu}&=&0\label{gauge}.
\end{eqnarray}
The Eq. (\ref{wave eq}) is a wave equation, the plane wave solution can be expressed as
\begin{eqnarray}
\phi_{\mu\nu}&=&A\varepsilon_{\mu\nu}\cos k^\alpha x_\alpha,
\end{eqnarray}
in which $A$ is the amplitude of the GWs, $\varepsilon_{\mu\nu}$ is polarization tensor,
and $k^\alpha$ is gravitational wave vector.
Wave vector $k^\alpha$ carries the information of the gravitational source,
from which the frequency, wavelength and propagating direction of the GWs  can be read.
From Eq. (\ref{gauge}) and Eq. (\ref{wave eq}), we have
\begin{eqnarray}
k^2&=&0,\\
k^\mu\varepsilon_{\mu\nu}&=&0.
\end{eqnarray}
Suppose the GWs propagates along the Z axis direction,
the gravitational wave vector $k^\alpha$ can be written as
\begin{eqnarray}
k^\alpha&=&(\omega_{\rm g},0,0,\omega_{\rm g}),
\end{eqnarray}
where $\omega_{\rm g}$ is the circular frequency of the GWs. The polarization tensor is
 some linear combination of vector $\varepsilon^\mu_1$, $\varepsilon^\mu_2$ and wave vector $k^\alpha$.
The vector  $\varepsilon^\mu_1$ and $\varepsilon^\mu_2$ is denoted as
\begin{eqnarray}
\varepsilon^\mu_1&=&(0,1,0,0),\\
\varepsilon^\mu_2&=&(0,0,1,0).
\end{eqnarray}
There are two physical polarization tensor $\varepsilon_{\mu\nu}$, it can be proven
that the rest of polarization tensors is just some gauge selection of $\phi_{\mu\nu}$.
Two transverse polarization tensor is
\begin{equation}
\varepsilon_{\oplus}^{\mu\nu}=\varepsilon^\mu_1\varepsilon^\nu_1-\varepsilon^\mu_2\varepsilon^\nu_2=
{\left(\begin{array}{cccc}
0 & 0 & 0 & 0\\
0 & 1 & 0 & 0\\
0 & 0 & -1 & 0\\
0 & 0 & 0 & 0
\end{array}
\right )},
\end{equation}
and
\begin{equation}
\varepsilon_{\otimes}^{\mu\nu}=\varepsilon^\mu_1\varepsilon^\nu_2+\varepsilon^\mu_1\varepsilon^\nu_2=
{\left(\begin{array}{cccc}
0 & 0 & 0 & 0\\
0 & 0 & 1 & 0\\
0 & 1 & 0 & 0\\
0 & 0 & 0 & 0
\end{array}
\right )}.
\end{equation}
The physical detection of these two kinds of polarization is similar.\cite{Riles:2012yw}
Thus in the following, we only study polarization tenser $\varepsilon_{\oplus}^{\mu\nu}$ .
The corresponding metric $g_{\mu\nu}$ is
\begin{equation}
g_{\mu\nu}=
{\left(
\begin{array}{cccc}
1 & 0 & 0 & 0\\
0 & h-1 & 0 & 0\\
0 & 0 & -h-1 & 0\\
0 & 0 & 0 & -1
\end{array}
\right )}.\label{h}
\end{equation}
Where
\begin{eqnarray}
h(x)=A\cos\big[\omega_{\rm g}\left(t-z\right)+\eta\big].
\end{eqnarray}
$\eta$ is the initial phase of the GWs.

From above formulations, we can see that the GWs propagating in Z axis direction
does cause the expansion or contraction in X or Y direction
periodically. Geometrically speaking,
the displacements in the X, Y direction is $dx, dy$ respectively,
the metric (\ref{h}) leads to the length change $X, Y$ in the form of
\begin{eqnarray}
dX&=&\sqrt{1-h}dx\simeq(1-\frac{1}{2}h)dx,\label{x}\\
dY&=&\sqrt{1+h}dy\simeq(1+\frac{1}{2}h)dy.\label{y}
\end{eqnarray}
Note that in the following section we will use the capital symbol $X$, $Y$ to denote
the proper and local length measured in the experiments.
As the proper length is invariant in the general relativity, the expansion and
the contraction here should take place in the coordinate. Or else, one can consider that
the ruler expands or contracts accompanying with the objects it measures.
The GWs in fact cause the oscillation of the coordinate. With the displacements above,
the next question is,
does the wavelength of laser in X and Y direction expands or contracts in the same way?
If so, then the optical path will be invariant no matter what kinds of metric is considered.
However, the wavelength of the laser is not an ordinary ruler, it is derived from
electromagnetic wave equation of laser, namely, from  the Maxwell equations.
Only the solver of the wave equation in the curved space-time metric (\ref{h}),
can  tell us how the electromagnetic waves change. Before that, let's take a
quick review on the theoretical principle of the GWs' detection by the
LIGO experiment.

\begin{figure*}
\centering
\scalebox{0.8}{\epsfig{file=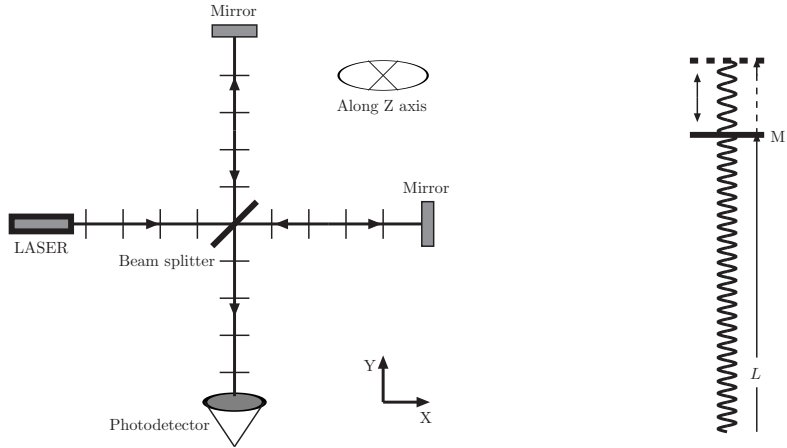}}
\caption{The left plot shows the detection of LIGO experiment,
the short thin line represents the polarization plane is parallel to XOY plane.
Any small displacement in the two arms (shown in the left plot) will change of the
interference pattern, output an optical signal to the detector. }
\label{fig2}
\end{figure*}

In the flat Minkowskian space time, the wave equation of electromagnetic waves is \cite{Jackson}
\begin{eqnarray}
\eta^{\mu\nu}\partial_\mu\partial_\nu F_{\alpha\beta}&=&0,\label{flatwave}
\end{eqnarray}
in which
\begin{equation}
F_{\alpha\beta}=
{\left(
\begin{array}{cccc}
0 & E_x & E_y & E_z\\
-E_x & 0 & -B_z & B_y\\
-E_y & B_z & 0 & -B_x\\
-E_z & -B_y & B_x & 0
\end{array}
\right )}
\end{equation}
is the electromagnetic field strength tensor.
The electric part of the plane wave solution  is
\begin{eqnarray}
&&E_x\sim\exp [i(\omega_{\rm em}t-k_{\rm em}y+\phi_1)],\label{E1}\\
&&E_y\sim\exp [i(\omega_{\rm em}t-k_{\rm em}x+\phi_2)],\label{E2}
\end{eqnarray}
where $\phi_{1,2}$  is the wave phase which can be adjusted by the laser.
In the experiment of the Michelson interferometer, which sketch graph of the
structure and measuring principle is shown in the FIG. \ref{fig2},
two polarized laser beams  travel along two  different path, arms along X axis and Y axis
respectively, and  recombine at beam-splitter.
The electromagnetic fields will interfere into different patterns.
By some adjustment in the wave phase of the initial beam,
the total light intensity received by photodetector is
\begin{eqnarray}
P&\simeq&({\rm Re}[\bm{E_x}]+{\rm Re}[\bm{E_y}])^2\nonumber\\
&=& P_0\sin^2[k_{\rm em}(X-Y)]\nonumber\\
&=&\frac{P_0}{2}\big[1-\cos(\Delta\varphi_{\rm em})\big].\label{basic}
\end{eqnarray}
In the above equation, $X$, $Y$ are the lengths of the two arms,
and $\Delta\varphi_{\rm em}$ is the phase shift of electromagnetic wave.
Therefore, by measuring  the light intensity, we can get the phase shift
 $\Delta\varphi_{\rm em}$ and thus get the the length differential between the two arms.
 Note that in case of  $X=Y$, the phase shift is canceled out at beam-splitter.

The above theoretical principle of the Michelson interferometer is actually
the detection principle of the most detection experiments of the GWs.\cite{Forward:1978zm}
Any change of optical path in Eq. (\ref{basic}) will cause a phase shift and create a signal.
LIGO detects the small displacement of arms influenced by the GWs.
By assuming that the GWs will influence the distance of the two arms,
and also, the propagation of the photons are the same as in the flat space time.
Using the Eq. (\ref{x}) and Eq. (\ref{y}), the phase shift $\Delta\varphi_{\rm em}$
can be approximately expressed as
\begin{eqnarray}
\Delta\varphi_{\rm em}&\simeq &k_{\rm em}hL,\label{ligoph}
\end{eqnarray}
where the total travelling distance of laser light is $L$, and the amplitude of
the GWs  $h$ can be obtained from the  phase shift $\Delta\varphi_{\rm em}$.
There may be a misunderstanding here that $hL$ is the difference between the
proper length of the arms. However, as talked above, $hL$ are the difference
of the coordinates which are oscillating when the GWs passing by.
Thus the phase shift in Eq. (\ref{ligoph}) is obtained under the
assumption that the velocity of phase transition in the coordinate space is invariant.
In case of the low frequency (kHz) GWs,
the LIGO experiment measured that\cite{Abbott:2016blz}
\begin{eqnarray}
h \sim 10^{-21}.
\end{eqnarray}
Then the phase shift is proportional to the strength of the GWs.
The amplitude $A$ is a very small quantity which is indirectly observed by LIGO detectors, in deed.
The shape change of two 4-kilometers arms is expected to have the
form of cosine function as Eq. (\ref{basic}).
Note that this is the case for the monochromatic GWs,
the wave equation is a simple harmonic oscillator.
However, if the GWs are consisted of many different frequency bands,
meaning that actually the gravitational wave packet is travelling,
then the wave equation of the GWs is the synthesis of many different  harmonic oscillator.
A gravitational wave packet could be possibly induced by its coupling with the
energy-momentum tensor $T^{\mu\nu}$ of material fields when the GWs are travelling.

As talked above, what we concern is the influence of the GWs on the laser propagation.
The shape change of the arms is not an observable quantity when being locally measured.
To figure out the answer, we will focus on the laser under the background GWs,
the Maxwell equations, wave equation and its solution, in the next section.
We should also note that the metric Eq. (\ref{h}) corresponds to a curved space-time.
Since the amplitude  $A$ is a much more small quantity,
any terms containing higher orders of $A$ are always neglected in the following calculation.
Moreover, if the electromagnetic field is strong enough, according to the Einstein equation,
the energy-momentum tensor of electromagnetic field $T_{\mu\nu}$ will also cause
a corresponding curved metric, this case is complicated and less concerned here, and it is
always neglected in the literature. In the following section,
we just focus on how the metric Eq. (\ref{h}) impacts on  electromagnetic waves.

\begin{figure*}
\centering
\scalebox{0.8}{\epsfig{file=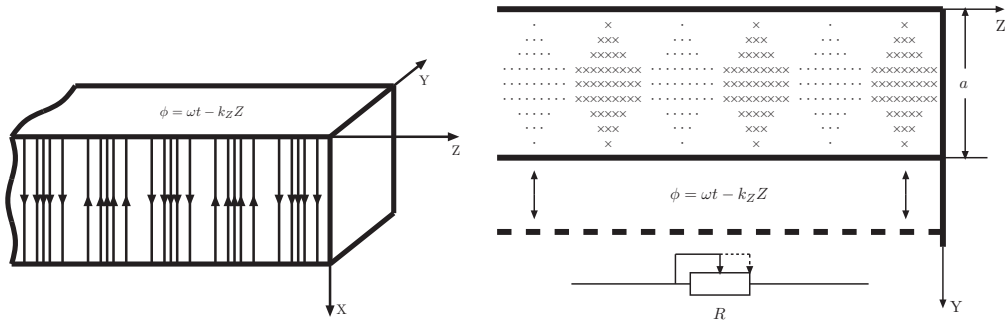}}
\caption{The ${\rm TE}_{01}$ wave mode in the waveguide.
The background of the GWs leads to dissipation and dispersion in the propagation of
electromagnetic wave. Like electromagnetic waves propagating in a waveguide with
resistance, as shown in the
figure, the phase is propagating along Z axis, forming the standing waves are along X and Y direction.
The phase of the GWs control the dissipation $c_1$ term  and dispersion $c_2$ term. }\label{fig3}
\end{figure*}
\section{Laser interference in curved space-time}\label{sec3}
According to the principle of general covariance,
generalizing a theory in the flat space to the curved space is simple,
what we should do is just to replace the ordinary derivative $\partial_\mu$
by the  covariant derivative $\nabla_\mu$ in ordinary Maxwell equations, i.e.
\begin{equation}
  \nabla_\mu A^\nu  = \partial_\mu A^\mu +\Gamma^\nu_{\mu\alpha} A^\alpha,
\end{equation}
where $\Gamma^\nu_{\mu\alpha}$ is the Christoffel symbol.
The form of the Maxwell equations with no source in curved space-time  are
\begin{eqnarray}
&& \nabla^{\mu}F_{\mu\nu}=0,\label{eem1}\\
&& \nabla_{\alpha}F_{\mu\nu}+\nabla_{\nu}F_{\alpha\mu}+\nabla_{\mu}F_{\nu\alpha}=0,\label{em2}
\end{eqnarray}
in which  $F^{\mu\nu}$ is  electromagnetic field tensor. \cite{Cabral:2016klm} 
Once again with covariant derivative $\nabla^\mu$ on Eq. (\ref{em2}), using  Eq. (\ref{eem1})
and relation
\begin{eqnarray}
\left(\nabla_{\beta}\nabla_{\alpha}-\nabla_{\alpha}\nabla_{\beta}\right)F_{\mu\nu}&=&
R_{\mu\alpha\beta}^{\rho}F_{\rho\nu}+R_{\nu\alpha\beta}^{\rho}F_{\mu\rho},
\end{eqnarray}
where $R^\alpha_{\beta\mu\nu}$ is Riemann curvature tensor,
we can obtain the wave equation of electromagnetic waves,\cite{Pfenning:2000zf}
\begin{eqnarray}
\nabla^{2}F_{\mu\nu}-R_{\mu}^{\rho}F_{\nu\rho}-R_{\nu}^{\rho}F_{\rho\mu}
+2R_{\mu\rho\nu\beta}F^{\rho\beta}&=&0.\label{emwaveeq}
\end{eqnarray}
$R_{\mu\nu}$ is Ricci tensor, defined as
\begin{eqnarray}
R_{\mu\nu}=R^\alpha_{\mu\nu\alpha}.
\end{eqnarray}
Note that here we use variable $F_{\mu\nu}$ rather than electromagnetic vector
potential $A^{\mu}$ in wave equations,
this is because that the polarization of laser can not  be expressed explicitly
when using  $A^{\mu}$. In the following we will find the polarization of
laser light is significantly linked to the GWs' detection.

At first we assume that the GWs propagate along the Z axis, and we check the
electromagnetic wave equation on the X and Y axis.
As the light is transverse wave, laser propagating in the X direction
only has y, z component of electromagnetic field,
and the one propagating in the Y direction only has x, z component, {\em et. al.}
Taking $\mu\nu=01$, the wave equation (\ref{emwaveeq}) gives
\begin{eqnarray}
&& g^{\mu\nu}\partial_\mu\partial_\nu F_{01}+A\omega_{\rm g}^{2}\cos[\omega_{\rm g}\left(t-z\right)]F_{01}\label{01}\\
&& -A\omega_{\rm g}\sin\left[\omega_{\rm g}\left(t-z\right)\right](F_{21,2}+F_{01,0}+
F_{03,1}-F_{01,3}) =0.\nonumber
\end{eqnarray}
Taking $\mu\nu=02$,
\begin{eqnarray}
&&g^{\mu\nu}\partial_\mu\partial_\nu F_{02}
+A\omega_{\rm g}^{2}\cos[\omega_{\rm g}\left(t-z\right)]F_{02}\label{02}\\
&&  +A\omega_{\rm g}\sin[\omega_{\rm g}\left(t-z\right)](F_{12,1}+F_{02,0}+
F_{03,2}-F_{02,3})=0.\nonumber
\end{eqnarray}
Taking $\mu\nu=03$,
\begin{eqnarray}
&&g^{\mu\nu}\partial_\mu\partial_\nu F_{03}\label{03}\\
&& +A\omega_{\rm g}\sin[\omega_{\rm g}\left(t-z\right)](F_{13,1}-F_{23,2}
+F_{01,1}-F_{02,2})=0.\nonumber
\end{eqnarray}
Note that all the  $A^2$ terms and higher orders of the metric in Christoffel
symbols $\Gamma^\alpha_{\mu\nu}$ and Ricci tensor $R_{\mu\nu}$  are neglected.

From wave Eqs. (\ref{01})-(\ref{03}), we can see that
the electromagnetic wave has a strong dependency  both on the
amplitude and the frequency of the GWs.  However, if we neglect the frequency $\omega_{\rm g}$
terms at first, we can get the answer for the above section, that is
whether the laser ruler expands or contracts in the same way as the length of the arms?
The equations of the plane electromagnetic wave shown in the FIG. \ref{fig1} should be
\begin{eqnarray}
&&  \left[\frac{\partial^2}{\partial t^2}-\frac{1}{1-h}\frac{\partial^2}{\partial x^2}\right]E_y=0,\label{eqx}\\
&&  \left[\frac{\partial^2}{\partial t^2}-\frac{1}{1+h}\frac{\partial^2}{\partial y^2}\right]E_x=0.\label{eqy}
\end{eqnarray}
The first equation is the wave equation along the X arms, and another one is for the Y arms. One
can see that the velocity of the wave phase can be faster or slower than the
velocity of light in the vacuum in the coordinate space.
However, this always happens in the general relativity
theory in which the coordinates may be not physical. The circumstance is contrary
to the assumption of universal velocity in the coordinate space
which is  talked in the above section. Thus the phase shift should not
be calculated by the difference between the length in the coordinates space.
In fact, these equations show that the laser ruler, namely the wavelength,
expands or contracts in the same way as the two arms of the interferometer. The
local measurement can not tell the variation of the metric.
To show it clearly, we can use the local physical proper distance $X$ and $Y$
\begin{eqnarray}
X&=&\sqrt{1-h}x,\\
Y&=&\sqrt{1+h}y.
\end{eqnarray}
Then the wave equation in the physical distance and time is
\begin{eqnarray}
&&\left[\frac{\partial^2}{\partial t^2}-\frac{\partial^2}{\partial X^2}\right]E_y=0,\label{phywx}\\
&&\left[\frac{\partial^2}{\partial t^2}-\frac{\partial^2}{\partial Y^2}\right]E_x=0.\label{phywy}
\end{eqnarray}
Eqs. (\ref{phywx})(\ref{phywy}) are exactly the same as Eq. (\ref{flatwave}).
the electromagnetic wave equation derived by the Maxwell equations in the vacuum
of the flat space. In fact, the local proper length of the arm is a physical quantity.
The wave equations (\ref{phywx})(\ref{phywy}) show the invariance of the physical velocity of
light which is the principle of the relativity. This means that using the apparatus and
the linearly polarized plane wave shown in the FIG. \ref{fig1} can not measure
the variation of the arms induced by the metric.
It seems that only the local measurement of effects of the extra $\omega_{\rm g}$ terms in
the Eqs. (\ref{01})-(\ref{03})
can show the difference between the arms and the ruler.  This is a little subtle  work,
we need  use some requirements on the variables appearing in the equations to
find the solution of the wave equations and  realize the real measurement.

The wave equations of electromagnetic waves consist of six partial differential
equations of $F_{\mu\nu}$.  The transverse laser implies that some components
of $F_{\mu\nu}$ absent in specific propagation direction.
For laser propagating in Y direction, we can chose $F_{01}$ and $F_{12}$
for consideration. Thus we require that
\begin{eqnarray}
F_{02} &=& F_{31} = F_{03}=0,\label{req1}\\
F_{21,2}&=&F_{01,0}-F_{31,3}+{\cal O} (A),\label{req2}\\
F_{01,3}&=&F_{03,1}+F_{31,0}+{\cal O} (A),\label{req3}
\end{eqnarray}
In which we omit all the higher orders terms in Eqs. (\ref{01})-(\ref{03}).
The resulted two Eqs. (\ref{req2})(\ref{req3}) are in fact the components
of the Maxwell equations.
These requirement are based on two reasons: first, linear polarization of the plane wave,
second, $A^2$ terms and higher terms in Eqs. (\ref{01})-(\ref{03})
are subleading terms. Under these requirements, the wave equations of
$F_{01}$, which describe the linear polarization
laser propagating in Y direction, now has the form of
\begin{eqnarray}
(\partial^{2}-2c_1\partial_t+c_2)E_x&=&0.\label{1F01}
\end{eqnarray}
Similarly, the wave equation propagation along X axis is
\begin{eqnarray}
(\partial^{2}+2c_1\partial_t+c_2)E_y&=&0.\label{1F02}
\end{eqnarray}
For $E_z$ along X or Y axis, the equation is
\begin{eqnarray}
(\partial^{2}-c_1\partial_t)E_z&=&0.\label{1F03}
\end{eqnarray}
In the above equations,  we denote
\begin{eqnarray}
c_1&\equiv&A\omega_{\rm g}\sin\left(\omega_{\rm g}t+\eta\right),\\
c_2&\equiv&A\omega_{\rm g}^{2}\cos(\omega_{\rm g}t+\eta).
\end{eqnarray}
We also assumed that $z=0$. These two terms $c_1$, $c_2$ are variables which
depends on the amplitude $A$ and circular frequency $\omega_{\rm g}$ of the GWs.
They are much smaller compared with the frequency of electromagnetic waves
in the experiment, Thus they can be considered as a constant in the wave equations.
We can see that there is no $c_2$ term and $c_1$ term lacks a factor 2
in Eq. (\ref{1F03}). This is because that the polarization is
along the direction of the GWs, the special direction in the circumstance.

Wave equations (\ref{1F01}) and (\ref{1F02}) are typical telegrapher's equations,
in which  the $c_1$ term is called the dissipation term, and the $c_2$ term is the dispersion term.
Usually the  telegrapher's equation describes the voltage and current on an
electrical transmission line with distance and time \cite{TE}, as shown in the FIG. \ref{fig3}
in which we show the ${\rm TE}_{01}$ mode in the waveguide. Note that
the phase velocity in the waveguide can always exceed the velocity of light in the vacuum
in the steady state.  Wave equations  (\ref{1F01}) and (\ref{1F02})
describe the influence of a passing GWs on the
propagation of  electromagnetic waves. Like electromagnetic waves in a waveguide with resistance,
the phase is propagating along Z axis, forming the standing waves are along X and Y direction.
The width of the Y boundary  of the waveguide is $a$, which determines the
the wavelength of the standing wave along Y axis. In the case of the real waveguide,
dispersion coefficient $c_2$ is determined by the widths of the waveguide,
i. e. ${\rm TE}_{mn}$ mode
\begin{equation}
c_2  = \frac{m^2\pi^2}{a^2 }+\frac{n^2\pi^2}{b^2 },
\end{equation}
in which $b$ is the width of the X boundary. In the background of the GWs,
the phase of the GWs control the dissipation $c_1$ term  and dispersion $c_2$ term.
However, as there is no boundary of the laser, $c_2$ can not be considered as coming from
the widths of a waveguide. There should be some additional parameters which
is determined by the laser. We will talk about them in the following.
If there is no GWs,  wave equations will return to the ordinary electromagnetic wave equation.
The dissipation term, arising from the  Christoffel symbols $\Gamma^\alpha_{\mu\nu}$,
causing an effective GWs-frequency-dependent resistivity which acts like a sliding resistance
as shown in the right plot of the FIG. \ref{fig3}.
The dispersion term, arising from the second derivative of metric,
provides an effective square of wave number $k^i_{\rm em}$ which determine the phase transition
along Z axis.

As there are dissipation and dispersion in the wave propagation, we can simply propose
a steady-state solution in the ${\rm TE}_{01}$ mode for the waves along different arms, i.e.
$E_x(t,Y)$ along the Y arm is
\begin{eqnarray}
E_x(t,Y)&=&E\exp(-\delta_1 Y)\sin(p_1 Z)\label{cos01p}
\\ && \times \exp{i(\omega_{\rm em}t-\beta_1 Y+\phi_1)},\nonumber
\end{eqnarray}
$E_y(t,X)$ along the X arm is
\begin{eqnarray}
E_y(t,X)&=&E\exp(-\delta_1 X)\sin(p_2 Z)\label{cos02p}\\
&& \times \exp{i(\omega_{\rm em}t-\beta_2 X+\phi_2)}.\nonumber
\end{eqnarray}
In above equations $p_{1, 2}$ are the additional parameters which is determined by the laser.
In the ordinary telegrapher's equation
\begin{equation}
  p_i^2 = c_2.
\end{equation}
As talked above, this condition is not necessarily hold in the background of the GWs. For the
simplicity, In the following, we consider the laser is plane wave which implies that
the variation of the field along Z direction can be neglected. In this case, Eqs. (\ref{cos01p})
(\ref{cos02p}) can be rewritten as
\begin{eqnarray}
E_x(t,Y)=E\exp(-\delta_1 Y)\exp{i(\omega_{\rm em}t-\beta_1 Y+\phi_1)},\label{cos01}&&\\
E_y(t,X)=E\exp(-\delta_2 X)\exp{i(\omega_{\rm em}t-\beta_2 X+\phi_2)}.\label{cos02} &&
\end{eqnarray}
Similarly $E_z(t,X^i_\perp)$ can be written as
\begin{eqnarray}
\hspace{-3mm}E_z(t,X^i_\perp)=E\exp(-\delta^\prime X^i_\perp)\exp{i(\omega_{\rm em}t
-\beta^\prime  X^i_\perp+\phi_3)}.\label{cos03}
\end{eqnarray}
In above equation,  $X^i_\perp$ means the directions perpendicular
to the propagation direction. $X^i_\perp$ can be $X$ or $Y$.
$\phi_{1,2,3}$ are the phase parameters, $\beta_{1,2}, \beta^\prime$  are the wave numbers.
In the language of waveguide physics, $1/\delta_{1,2}, 1/\delta^\prime$ is the skin depth,
they cause an exponential depression of the amplitude $E$ along propagating axis.
Eqs. (\ref{cos01})-(\ref{cos03}) can also have an increasing form
of the amplitude which we will not discuss in the context.
We assume that the intensity of the photon flux is dominantly
controlled by the phase of the electromagnetic waves.
Solving the Eqs. (\ref{1F01})-(\ref{1F03}), we can get
\begin{eqnarray}
 \delta_1&=& \delta_2 =\frac{\omega_{\rm em}c_1}{\beta},\label{delta}\\
 \beta_1^2&=& \beta_2^2 = \beta^2 \nonumber \\
&=&\frac{\omega_{\rm em}^2-c_2
+\sqrt{(\omega_{\rm em}^2-c_2)^2+4\omega_{\rm em}^2c_1^2}}{2},\label{beta}\\
\delta^\prime&=&\frac{\omega_{\rm em}c_1}{2\beta^\prime},\label{deltap}\\
\beta^{\prime2}&=&\frac{\omega_{\rm em}^2+\sqrt{\omega_{\rm em}^4
+\omega_{\rm em}^2c_1^2}}{2}\label{betaprime}.
\end{eqnarray}
We can see that $c_1$ and $c_2$ which are from the parameter set of the GWs
determine the wave vector $k^\mu_{\rm em}$.
Note that $c_1$ terms in $\beta$ and $\beta^\prime$ always appear in quadratic form,
which can be ignored in the following estimation of the phase propagation.
This means that the differences between the phase velocity
dominantly comes from the $c_2$ term.
While wave equation of $E_z$ of Eq. (\ref{cos03}) can be distinguished from $E_x$ and $E_y$.
through a different wave number $\beta^\prime$ and a factor $2$ in $c_1$ term.
The phase velocity of $E_x$, $E_y$ and $E_z$ is
\begin{eqnarray}
v_x&=&\frac{\omega_{\rm em}}{\beta},\label{phvx}\\
v_y&=& v_x, \label{phvy}\\
v_\perp&=&\frac{\omega_{\rm em}}{\beta^\prime}\label{phvz}.
\end{eqnarray}
The plane wave solution  of Eq. (\ref{cos01})
and Eq. (\ref{cos02}) has the same wave phase velocity.
As the interference pattern dominantly determined by the phase, we
can  ignoring the dissipation term including $\delta$ and $\delta^\prime$.
The total light intensity $P$ for the plane wave $E_x$, $E_y$ is
\begin{eqnarray}
P&\simeq&(E_x+E_y)^2= P_0\sin^2\big[\beta(X-Y)\big].\label{tp}
\end{eqnarray}
Since the physical length of two arms are invariant, and the phase velocity is the same too,
if the propagation wave along X and Y arms obey these two equations, there will be
no changing of interference pattern. This case is similar to the case
that ignoring the linear $A$ terms of the wave equation,
as talked above. Two linearly polarized laser beams interfere mutually.
The same pattern of expansion and contraction leads to an
invariant proper distance, eliminating any signal of the
 GWs. The same form of electromagnetic wave number is another factor leading to this situation.
Neither the dissipation term nor the dispersion term  contribute to the total light intensity formula.
In the case of the invariant proper distance, only different wave number of
two laser light can cause a phase shift. Such circumstance could happen
in the consideration of a specific  linearly polarized laser.
From the Eqs. (\ref{phvx})-(\ref{phvz}), we can see that
interference of $E_z$ with  $E_x$ or $E_y$
can give a local measurement on the metric.

\begin{figure*}
\centering
\scalebox{0.8}{\epsfig{file=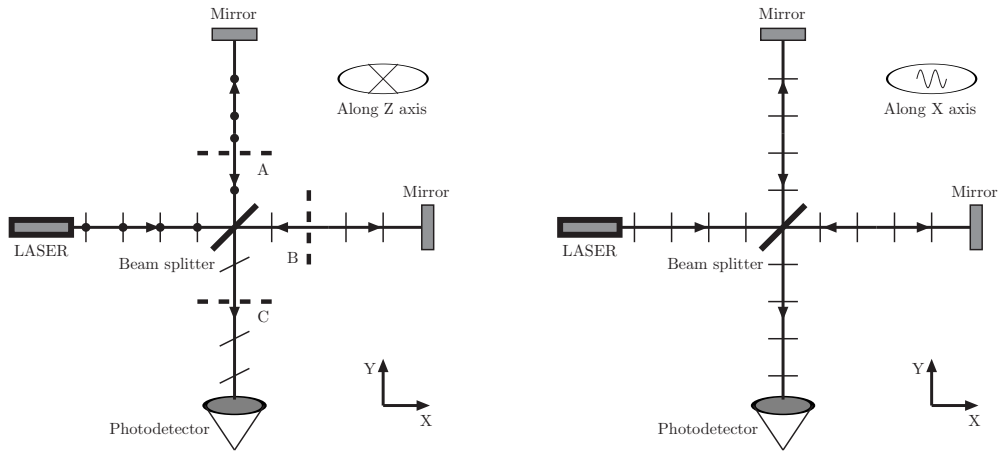}}
\caption{Interference of electromagnetic waves in the Michelson interferometer.
The left panel: the GWs propagate along Z axis, wave which emits from the laser is
unpolarized, and A, B and C are Polaroids. A (B) makes the laser along Y (X) axis
polarized along Z (Y) axis. C makes the two mutually interference of the two beams
polarized to a direction with included angle equals to $45^\circ$ between the two beams.
The right panel: the GWs propagate along X axis, wave which emits from the laser is
polarized along the XOY plane can interfere. The change of the interference pattern
give a signal of the GWs.}
\label{fig4}
\end{figure*}
Take $E_x$ and $E_z$ interference as a example. Despite of the invariant proper distance,
different wave number will generate a phase shift, leading to an ideal optical signal.
However, the polarization directions of $E_x$ and $E_z$ are perpendicular to each other,
we need additional apparatus to realize the interference.
In fact, in the interference there is a special direction: the direction of the GWs.
Thus, any change of the interference pattern should involve in this special direction.
For example, we  can use  a specific  polaroid to change the polarization
direction of the laser before arriving at photodetector.
An optical signal could possibly be detected, indicating
a passing GWs propagating in Z axis direction.
Or else, We need  one linearly polarized laser
beam propagating  in the direction of the incoming GWs to give
a different wave number compared with another a paralleled polarized plane
wave. This could be realized by disposing an arm along Z axis
with another one unchanged.
In this situation, an interference pattern could be formed
at photodetector without a crystal.
The detail will be discussed in the following section.

\section{Inference on the gravitational waves and comparison with the
current detection results}\label{sec4}
In this section, we show the phase shift of our result and the implication on the
GWs. Also we will discuss and compare the result with the detection results in the market.
First we should make clear about the difference between the detection
of the GWs in the above two section. All the detection are based on the change
of the interference pattern which gives the phase shift of electromagnetic wave
in the detector of interferometer. In the sec. \ref{sec2}, the phase shift comes from the
changing of the difference between the lengths of the two arms in the coordinate space.
The velocity in the coordinate space is invariant. In the sec. \ref{sec3}, we
calculate the electromagnetic waves in the curved space, finding that the length of the
wave expands and contracts accompanying with the arms. It's hard to detect
the difference of length of the arms. Only special interference mode
can give proper phase shift. We think
only these solutions can give a more robust principle of the detection
of the GWs. Thus in the following we show the detail to infer the amplitude and frequency of
the GWs in the interference of the laser.

As talked in the above section, only the interference between the laser along the direction of the GWs
and along other direction can give proper local measurement of the GWs. Thus
the modification of interference is illustrated in the left plot of Fig \ref{fig4},
in which two polaroids A and B generate two orthogonal linearly polarized plane laser beams.
A special  crystal C makes the two beams polarized to a direction
with included angle equals to $45^\circ$ between them, forming an interference pattern.
However, there is no design of these apparatus equipped in the detectors
of the GWs in the market. Thus, the only left possibility of the
signal of the GWs detected on the interferometer
is that the GWs must propagate along one arm of the interferometer. Say, along the X
axis, which is  shown in the right panel of the FIG. \ref{fig4}.
In this case, a phase shift could be generated without
additional polaroids. Thus we will take this situation to check
the phase shift measured by the LIGO collaboration.

As for GWs propagating in X axis direction, the laser propagating in
Y axis direction only has $E_x$ component, laser propagating in X axis direction
only has $E_y$ component. In this circumstance, the wave equation along Y axis
will change to the Eq. (\ref{cos03}) with corresponding change of the $E_z$ to $E_x$.
However, the wave equation along the X axis is not exactly the telegrapher's equation. It
has to change to
\begin{equation}
  \left[\partial^2 +c_1\partial_t - c_1\frac{\partial}{\partial X} + c_2\right]E_y = 0,
\end{equation}
according Eq. (\ref{02}). The solution of the equation is similar, we can assume that
\begin{eqnarray}
\hspace{-0.7cm}E_x(t,Y)&=&E\exp(-\delta^\prime Y)\exp{i(\omega_{\rm em}t-\beta^\prime Y+\phi_1)},\\
\hspace{-0.7cm}E_y(t,X)&=&E\exp(-\delta X)\exp{i(\omega_{\rm em}t-\beta X+\phi_2)}.
\end{eqnarray}
The result of $\delta^\prime$ and $\beta^\prime$ are the same as Eqs. (\ref{deltap})(\ref{betaprime}).
$\delta$ and $\beta$ become much more complicated, i. e.
\begin{equation}
  \delta = \frac{(\omega_{\rm em}+\beta)c_{1}}{2\beta}.
\end{equation}
To get the dispersion relation one needs to solve the following equation
\begin{equation}
 \omega_{\rm em}^{2}+\frac{(\omega_{\rm em}+\beta)^2c_{1}^2}{4\beta^2}-\beta^{2}-
\frac{(\omega_{\rm em}+\beta)c_{1}^2}{2\beta}-c_{2}=0.
\end{equation}
The analytical expression of $\beta$ is complicated, However, as talked above,
the $c_1^2$ term can be neglected, the phase difference dominantly comes from $c_2$ terms.
Thus the analytical $\beta$ is not shown here for its little influence on the phase shift.
The total light intensity formula will be
\begin{eqnarray}
P\simeq P_0\sin^2(\beta^\prime X-\beta Y)
=\frac{P_0}{2}\big[1-\cos(\Delta\beta L)\big]
\end{eqnarray}
The differences in wave number $\Delta \beta=\beta^\prime-\beta$
could cause a phase shift and generate an optical signal.
Note that, in the ordinary detection principle talked in the sec. \ref{sec2}, the phase shift
comes from $\Delta L$, Here the phase shift comes from the change of the wave number.
In SI unit, use the formulas (\ref{beta}) and (\ref{betaprime}) up to the first order of $A$,
\begin{eqnarray}
\Delta \beta(A)&\simeq& aA,
\end{eqnarray}
in which
\begin{eqnarray}
a&=&\frac{\omega_{\rm g}^{2}\cos(\omega_{\rm g} t+\eta)}{2\omega_{\rm em}c_{\rm g}}.
\end{eqnarray}
The dimension of $a$ is  $L^{-1}$, and $c_g$ is the speed of GWs and
we assume $c_g$ equals to the velocity of light in the vacuum.\cite{TheLIGOScientific:2017qsa}
We can see that there is  a suppression factor $\omega_{\rm g}/\omega_{\rm em}$,
which implies that if the frequency of the electromagnetic wave is much greater
than the frequency of the GWs, $a$ will be highly suppressed by the factor.
This is different from the ordinary detection principle in which the frequency
of the GWs is absent. The total phase shift can be obtained
\begin{eqnarray}
\Delta\varphi_{\rm em}=\Delta \beta L.
\end{eqnarray}
We use the  phase shift detected by LIGO detectors, (Which we calculated
from Eq. (\ref{ligoph}) by using the $h$ measured by LIGO.)
we can obtain the amplitude $A$ by
\begin{eqnarray}
A&=&\frac{\Delta \varphi_{\rm em}}{a L}.
\end{eqnarray}
If we chose the frequency of the GWs at $100$Hz and the
frequency of the laser is $10^{14}$Hz, then we can get the
amplitude $A$ will be order 1 which is much greater than
the amplitude assumed in the community.\cite{Riles:2012yw}
Is our consideration and
calculation wrong? We don't think so. The case is much complicate than
the usual one. Detection of the amplitude $A$ depends on the frequency of the GWs and
the propagation of the electromagnetic waves. If the GWs contain some
high frequency modes, say $\nu_{\rm g}\sim 10^9$Hz,
the corresponding amplitude will be
\begin{eqnarray}
A&\sim&6.3\times10^{-12},
\end{eqnarray}
which is much comfortable for the audience. In fact, such high frequency modes
are absolutely needed in the detection of the GWs, which we will show in the
following.

Here we give a summary about what we are doing now. In sec. \ref{sec3} we find that
a local measurement of the metric is impossible if we ignore the differential
of the metric. In this section, we chose the special direction of the GWs and
make the measurement possible. We emphasis that the key point  in the
theoretical side is second derivative of metric in Eq. (\ref{emwaveeq})
gives the dominant $c_2$ terms. $c_1$ terms which are the first
derivative of metric are subleading terms. While the absolute  value of metric
are in fact irrelevant in local measurement.

\begin{figure*}
\centering
\scalebox{0.8}{\epsfig{file=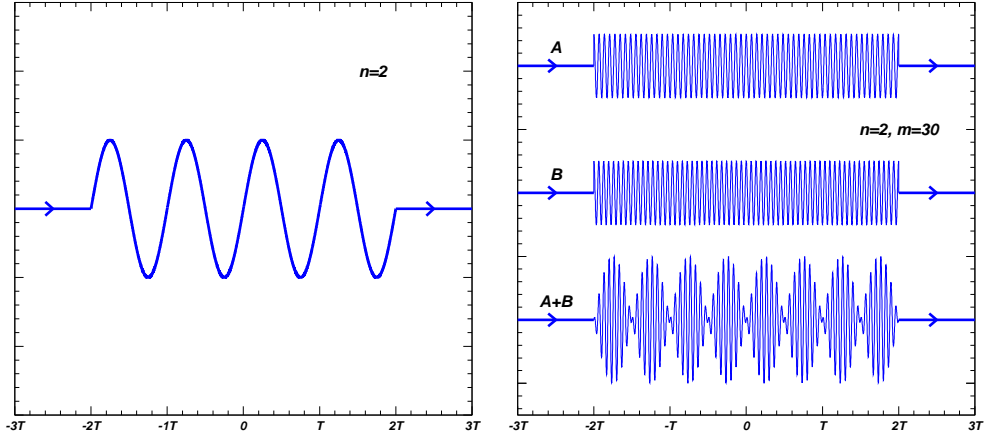}}
\caption{A pulsed monochromatic wave and a pulsed beats of the GWs: the left plot
 is pulsed monochromatic wave with $2\times2$ periods and the right one
is a pulsed beats which frequency equals to the monochromatic frequency of the left plot.
There are $m=30$ periods during one beat.}
\label{fig5}
\end{figure*}

All the discussions in above section are in assumption that a
plane gravitational wave passing through the interferometer. However,  the
signals of the GWs are all in fact pulsed signals which last less than one second.
If we are serious about this, at first, the  GWs seems to be a pulsed
wave which is shown in the left panel of FIG. \ref{fig5}. If we use the
Eqs. (\ref{1F01})-(\ref{1F03}) to research the interference,
we need to Fourier expansion of $h(t)$  to get amplitude of every plane wave modes
$H(\omega_{\rm g})$ and calculate the corresponding  phase shift
$\Delta \varphi(\omega_{\rm g})$.  The convolution of $\Delta \varphi(\omega_{\rm g})$ in the
$\omega_{\rm g}$ space can give the final phase shift to be measured. Nevertheless,
every interference pattern are in fact a steady modes which results from an
assumption that the interference relaxation time of electromagnetic wave is
much shorter than the period of the GWs. This condition can not
hold in case of the appropriately high frequency of the GWs.

\begin{figure*}
\centering
\scalebox{0.8}{\epsfig{file=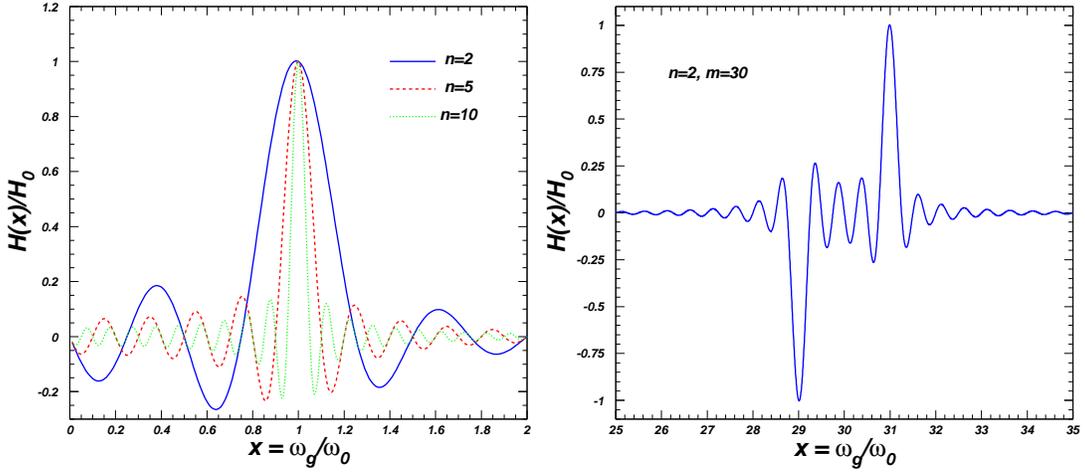}}
\caption{Same as FIG.\ref{fig5}, but shows the Fourier expansion of the
pulsed monochromatic wave (left) and pulsed beats (right).}\label{fig6}
\end{figure*}

For simplicity, we consider the GWs is a  wave packet composed of a monochromatic wave
\begin{eqnarray}
h(t)=A\sin\omega_{0}t,
\end{eqnarray}
in which $\omega_0$ is the monochromatic frequency,  and there
are  $2n$ period during the pulse which implies that the duration time $t$ is
\begin{eqnarray}
-n\frac{2\pi}{\omega_{0}}<t<n\frac{2\pi}{\omega_{0}}.
\end{eqnarray}
As talked in above paragraph, we need the corresponding
Fourier expansion $h(t)$  for the calculation of the phase shift,
\begin{eqnarray}
H(\omega_{\rm g}/\omega_0)&=& \frac{A}{2\pi}\int_{-\frac{2n\pi}{\omega_{0}}}^{\frac{2n\pi}{\omega_{0}}}\sin\omega_{0}t\sin\omega_{\rm g}t{\rm d}t\\
&=& H_0 f_1(\omega_{\rm g}/\omega_0)
\nonumber
\end{eqnarray}
in which
\begin{eqnarray}
  H_{0} = \frac{nA}{\omega_0},
\end{eqnarray}
and
\begin{eqnarray}
  f_1(x)= \frac{1}{2n\pi}\biggl[\frac{\sin[(1-x)2n\pi]}{1-x}-\frac{\sin[(1+x)2n\pi]}{1+x}\biggl].
\end{eqnarray}
The amplitude in the frequency space is shown in the in left panel
 of  FIG. \ref{fig6} in which we show   $n=2, 5, 10$, respectively.
We can see that though the  $\omega_{\rm g}$ has the largest amplitude at
$\omega_0$, there are sizable amplitudes at other frequencies, especially when
$n$ is smaller. However, they decays  quickly with the growth of
the deviation  from  the monochromatic frequency $\omega_0$.
We can imagine that the convolution of the $\Delta \varphi$ is still depends
on $\omega_0$, thus, low frequency of the GWs still implies that
the amplitude $A$ is incredibly large.

\begin{figure*}
\centering
\scalebox{0.6}{\epsfig{file=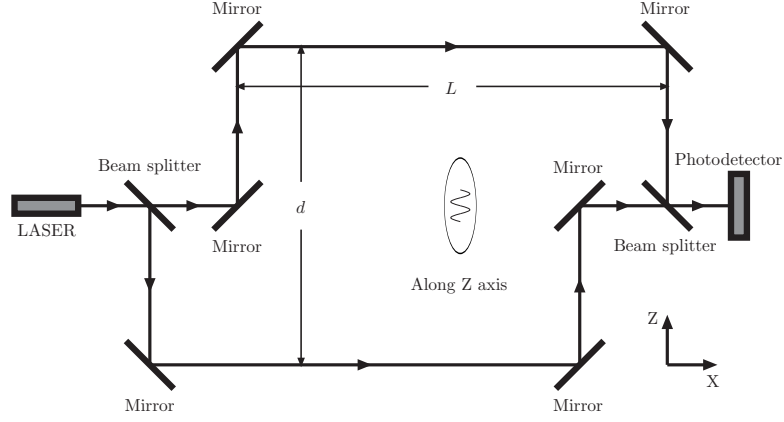}}
\caption{A proposal of a detection of the GWs with high frequency of beats.
the GWs propagating along Z axis, and the length between the two neighbour beats
is assumed to be $d$. The upper and the lower horizon lines expand of
contract when the GWs passing by. The no local interference can give change
of the interference pattern at the detector.}
\label{fig7}
\end{figure*}
What we will do now is to propose a pulsed beats which is a special phenomenon
in the waves to escape from the dilemma.  The synthesis of
two harmonic oscillations with different but almost equal frequency gives
the beats which shown in the right panel of the FIG. \ref{fig5}.\cite{berwaves}
The amplitude of the beats is not a constant but a modulation of almost harmonic
oscillation. Frequency of beats is much smaller than the frequency of
the harmonic oscillation. For example, we chose a beats as
\begin{eqnarray}
h(t)=A\sin\omega_{0}t\cos m\omega_0t.
\end{eqnarray}
in which $A\sin\omega_{0}t$ is the amplitude of the modulation. For simplicity,
$m$ is an integer which can be much greater than $n$. If the beats is also
pulsed as shown in the right panel of FIG. (\ref{fig5}), in which we take
$n=2, m=30$ for example. Fortunately, we can get a much larger frequency
amplitude in the frequency space. The Fourier expansion $h(t)$ is similar
\begin{eqnarray}
H(\omega_{\rm g}/\omega_0)&=& \frac{A}{2\pi}
\int_{-\frac{2n\pi}{\omega_{0}}}^{\frac{2n\pi}{\omega_{0}}}\sin\omega_{0}t\cos m\omega_0t
\sin\omega_{\rm g}t{\rm d}t\nonumber \\
&=& H_0^\prime f_2(\omega_{\rm g}/\omega_0),
\end{eqnarray}
in which
\begin{eqnarray}
  H_0^\prime =\frac{nA}{2\omega_0},
\end{eqnarray}
and
\begin{eqnarray}
f_2(x)&& = \frac{1}{2n\pi}\\
&& \times \left[\frac{\sin[(m+1)-x]2n\pi}{[(m+1)-x]}
-\frac{\sin[(m+1)+x]2n\pi}{[(m+1)+x]}\right.\nonumber\\
&& \left. -\frac{\sin[(m-1)-x]2n\pi}{[(m-1)-x]}+\frac{\sin[(m-1)+x]2n\pi}{[(m-1)+x]}\right].\nonumber
\end{eqnarray}
Numerical results of $f_2(\omega_{\rm g}/\omega_0)$ are shown in the right panel of the FIG. \ref{fig6}.
We can see that frequency with maximum  amplitude can be much greater than the $\omega_0$.
This means that we can chose a pulsed beats with a much higher frequency of the almost harmonic
oscillation and very low frequency of beats. This may be a similar signal obtained by the LIGO.
Note that, the simple gravitational wave equation Eq. (\ref{wave eq}) are obtained under
the low  frequency $\omega_{\rm g}$ assumption. In case of a high frequency,
gravitational wave equation is not a simple linear wave equation. 
The non-linear wave equation and the solution are beyond this work. \cite{Culetu:2016eek} 
The  high frequency $\omega_{\rm g}$ implies a small wavelength of the GWs, thus the
calculation of the phase shift along the direction of the GWs may be not a local quantity.

However,  one may suspect on the assumption above that the detector only shows a pulsed beats.
As the oscillation of the metric is so fast, the interference pattern may be able to be detected.
We should argue that in case of the much faster harmonic oscillation, the relaxation of the
interference must be considered. This can be understood as following:  the laser beams  need
a relaxation time to form a steady-state, which can be evaluated by time interval of
 the laser propagating along the arms of the interferometer. For LIGO experiment the
relaxation time $\Delta t_r$
\begin{eqnarray}
\Delta t_r\simeq\frac{4\rm km}{c}\simeq10^{-5}s.
\end{eqnarray}
If the frequency of GWs is much higher, i.e. $\nu_{\rm g}=10^9$ Hz which has been mentioned
 above, the period of the GWs will be about of  $10^{-9}s$ which is much smaller
than the $\Delta t_r$, a signal will be disturbed before it forms into a steady-state.
Thus a clean signal in fact is not able to be detected.
Only the modulation of the pulsed beats is detected. The detail of
the detection of the relaxation and the pattern is beyond this work.
Note that, $\nu_{\rm g}=10^9$ Hz implies that the amplitude is still much greater
than the amplitude predicted by the numerical simulation of coalescence
of two black holes. In this work, we concentrate on the detection of  the GWs,
as for the source of the GWs, we argue that, coalescence of two black holes
may be much complicate than the ordinary simulation.\cite{Cao:2008wn}

In all, the key point of our paper is that it is very difficult to detect the variation of
the metric in a local measurement. Even we solve the Maxwell equation in the curved
space, we can see that, only introducing a much complicated pulsed beats can give
an appropriate explanation of the detection result of the GWs in the market.
Similar to that the phenomenon of redshift verifies the theory of the general relativity,
a globally comparison of length of time can easily verify the variation of the metric.
Thus we propose a non-local interference which is shown in the FIG. \ref{fig7}
to detect the high-frequency GWs. The setting of the apparatuses
is show in the figure. In the experiment,
the initial laser beam is divided into two beams by the beam-splitter and mirror,
the divided beams propagate in an opposite direction along the Z axis. After that
they are reflected to propagate along X axis (horizon line in the FIG. \ref{fig7} )
at different coordinate of Z axis.  The distance between them is $d$. The two beams
meets each other at the detector by  reflections.
Assume a GWs is propagating along the Z axis, the upper arm and the lower arm
have different wave phase. The changing of interference pattern
in the detector will give a global
measurement of  small difference between the lengthes of the upper and lower arms.
This is similar to the measurement of redshift caused by gravitation.

The best event of the GWs will be that the distance $d$ happens to
be equal to one half of the GWs' length. Then the difference
of the two arms will be largest in the duration of the GWs.
There is always a  phase difference of the GWs between two arms.
Suppose the gravitational phase of upper arm is $\eta_1$, and $\eta_2$ for lower arm.
To generate the phase shift of laser light, we consider the wave equation of $E_y$ and $E_y^\prime$,
for laser travelling along the upper and the lower arms respectively,
\begin{eqnarray}
\big[\partial^{2}+2c_1(\eta_1)\partial_t+c_2(\eta_1)\big]E_y&=&0,\\
\big[\partial^{2}+2c_1(\eta_2)\partial_t+c_2(\eta_2)\big]E_y^\prime&=&0.
\end{eqnarray}
Where $c_1(\eta_i)$ and $c_2(\eta_i)$ refer to
\begin{eqnarray}
c_1(\eta_i)&=&A\omega_{\rm g}\sin\left(\omega_{\rm g}t+\eta_i\right),\\
c_2(\eta_i)&=&A\omega_{\rm g}^{2}\cos(\omega_{\rm g}t+\eta_i).
\end{eqnarray}
These wave equations lead to the same expression of  wave number but
there is a phase difference between them
\begin{eqnarray}
\Delta \beta&=&\beta(\eta_2) - \beta(\eta_1).
\end{eqnarray}
Two beams of laser will interfere, the corresponding
phase shift can be obtained by the same expressions above.
Up to the first order of amplitude $A$,
\begin{eqnarray}
\Delta\varphi_{\rm em}&=&\Delta\beta L\\
&=&\frac{A\omega_{\rm g}^{2}L}{2\omega_{\rm em}c_{\rm g}}
\big[\cos(\omega_{\rm g} t+\eta_1)-\cos(\omega_{\rm g} t+\eta_2)\big].\nonumber
\end{eqnarray}
Similarly, this phase shift leads to an optical signal for photodetector.

We know that, the experiment we proposed above always suffers the space it
occupies. $d$ can not be to large, say about $3 \rm km$, implying that a high frequency of the
GWs. Thus the respondent time and the sensitivity of the detector will be a problem  in the
real experiment. However, similar circumstance can be detected too, that the experiment
can detect pulsed beats, which can give a global measurement of the beats.
Together with the ordinary detector as shown in the FIG. \ref{fig4}, they will give us
a further and mutually measurements of the GWs.

\section{Conclusion}\label{sec5}
In the work, we checked laser interference in the background of the GWs.
In the ordinary principle of the interference measurement,
the laser is assumed as a flux of photon
with a speed of light which is unaffected by the GWs. The measurement
is the variation of intensity which can give the  phase  shift of the electromagnetic waves,
and then the  length displacements of the arms in the interferometer.
However, we find that the expansion and contraction of wave lengths
are always synchronous with the arms. We must be very careful on the local
measurement of the variation of metric. Like electromagnetic waves propagating in a waveguide with
resistance, the background of the GWs leads to  dissipation and dispersion in the propagation of
electromagnetic wave. The phase of the GWs controls the
dissipation term  and dispersion  term in the telegrapher's equation.
In fact, in the interference of laser in the local measurement, there is
the special direction which is the direction of the GWs.
Any change of the interference pattern should involve in this special direction.

In order to detect the GWs, we  can use  a specific  polaroid to change the polarization
direction of the laser before it arriving at the photodetector. An optical signal could possibly
be detected, indicating a passing GWs. Or else, we need  one linearly polarized laser
beam propagating  in the direction of the incoming GWs to give
a different wave number compared with the linearly polarized plane
wave in another path. We use the phase shift measured by the LIGO to check the case of pulsed
plane wave, finding that, the amplitude is incredibly large. However, we
propose a pulsed beats to solve the problem. Our analysis show that, in case of
the pulsed beats passing by, the relaxation time is greater than the period of
the GWs, the signal will be disturbed before it forms a steady-state,
thus a clean signal in fact is not able to be detected. The detector may only show
a signal of the modulation of the beats.

We should comment that, in this paper, we emphasised that the local measurement can
not show the variation of metric. This is because a local distance and time interval must
be defined by the invariance of velocity of light in the vacuum. This is a
fundamental principle of relativity. Only global comparison can show the difference of the
metric. Nevertheless, Einstein equation of the gravitation is the dynamics of the metric,
Any physical quantity derived from Einstein equation must involve differential of
the metric, not only the absolute value. Thus what we should concern in the expansion or
the approximation of the gravitation are in fact the derivative of the metric.
As shown in the paper, its effects are really what we can measure in the local experiment.

Finally we proposed a non-local interference experiment
to detect the high-frequency GWs. It is similar to the measurement
of redshift caused by gravitation. Together with the ordinary detector,
it will give us further and mutually measurements of  the GWs. Of course
the detail of the relaxation of the interference pattern, {\em et. al.}
needs our further studies.

\section*{ACKNOWLEDGEMENT}
We thanks Bin Zhu, Wei Xu for very useful discussion and comment on our work.
This work was supported by the Natural Science Foundation of China under grant number 11775012.


\begin{thebibliography}{10}
\bibitem{TheLIGOScientific:2016agk}
B.~P.~Abbott {\it et al.} [LIGO Scientific and Virgo Collaborations],
%  ``GW150914: The Advanced LIGO Detectors in the Era of First Discoveries,''
  Phys.\ Rev.\ Lett.\  {\bf 116}, no. 13, 131103 (2016)

\bibitem{Vitale:2016rfr}
S.~Vitale,
%  ``Multiband Gravitational-Wave Astronomy: Parameter Estimation and Tests of General Relativity with Space- and Ground-Based Detectors,''
  Phys.\ Rev.\ Lett.\  {\bf 117}, no. 5, 051102 (2016)

\bibitem{Abbott:2016nmj}
 B.~P.~Abbott {\it et al.} [LIGO Scientific and Virgo Collaborations],
%  ``GW151226: Observation of Gravitational Waves from a 22-Solar-Mass Binary Black Hole Coalescence,''
  Phys.\ Rev.\ Lett.\  {\bf 116}, no. 24, 241103 (2016)

\bibitem{Abbott:2017vtc}
B.~P.~Abbott {\it et al.} [LIGO Scientific and VIRGO Collaborations],
%  ``GW170104: Observation of a 50-Solar-Mass Binary Black Hole Coalescence at Redshift 0.2,''
  Phys.\ Rev.\ Lett.\  {\bf 118}, no. 22, 221101 (2017)

\bibitem{Abbott:2017oio}
B.~P.~Abbott {\it et al.} [LIGO Scientific and Virgo Collaborations],
%  ``GW170814: A Three-Detector Observation of Gravitational Waves from a Binary Black Hole Coalescence,''
  Phys.\ Rev.\ Lett.\  {\bf 119}, no. 14, 141101 (2017)

\bibitem{Abbott:2017gyy}
B.~. P.~.Abbott {\it et al.} [LIGO Scientific and Virgo Collaborations],
%  ``GW170608: Observation of a 19-solar-mass Binary Black Hole Coalescence,''
  Astrophys.\ J.\  {\bf 851}, no. 2, L35 (2017)

\bibitem{TheLIGOScientific:2017qsa}
B.~P.~Abbott {\it et al.} [LIGO Scientific and Virgo Collaborations],
%  ``GW170817: Observation of Gravitational Waves from a Binary Neutron Star Inspiral,''
  Phys.\ Rev.\ Lett.\  {\bf 119}, no. 16, 161101 (2017)

%\cite{Cabral:2016klm}
\bibitem{Cabral:2016klm} 
  F.~Cabral and F.~S.~N.~Lobo,
  %``Gravitational waves and electrodynamics: New perspectives,''
  Eur.\ Phys.\ J.\ C {\bf 77}, no. 4, 237 (2017)
%  doi:10.1140/epjc/s10052-017-4791-z
  [arXiv:1603.08157 [gr-qc]].
  %%CITATION = doi:10.1140/epjc/s10052-017-4791-z;%%
  %4 citations counted in INSPIRE as of 09 Nov 2018

\bibitem{Gondan:2018khr}
  L.~Gond\'an and B.~Kocsis,
  arXiv:1809.00672 [astro-ph.HE].
  %%CITATION = ARXIV:1809.00672;%%
  %2 citations counted in INSPIRE as of 11 Nov 2018

\bibitem{Goldstein:2017mmi}
A.~Goldstein {\it et al.},
%  ``An Ordinary Short Gamma-Ray Burst with Extraordinary Implications: Fermi-GBM Detection of GRB 170817A,''
  Astrophys.\ J.\  {\bf 848}, no. 2, L14 (2017)

  \bibitem{Kramer:2013kea}
 M.~Kramer and D.~J.~Champion,
%  ``The European Pulsar Timing Array and the Large European Array for Pulsars,''
  Class.\ Quant.\ Grav.\  {\bf 30}, 224009 (2013).

\bibitem{McLaughlin:2013ira}
 M.~A.~McLaughlin,
%  ``The North American Nanohertz Observatory for Gravitational Waves,''
  Class.\ Quant.\ Grav.\  {\bf 30}, 224008 (2013).

\bibitem{Hobbs:2013aka}
G.~Hobbs,
%  ``The Parkes Pulsar Timing Array,''
  Class.\ Quant.\ Grav.\  {\bf 30}, 224007 (2013)


\bibitem{Harry:2010zz}
 G.~M.~Harry [LIGO Scientific Collaboration],
%  ``Advanced LIGO: The next generation of gravitational wave detectors,''
  Class.\ Quant.\ Grav.\  {\bf 27}, 084006 (2010).


\bibitem{TheVirgo:2014hva}
F.~Acernese {\it et al.} [VIRGO Collaboration],
%  ``Advanced Virgo: a second-generation interferometric gravitational wave detector,''
  Class.\ Quant.\ Grav.\  {\bf 32}, no. 2, 024001 (2015)

\bibitem{Aso:2013eba}
Y.~Aso {\it et al.} [KAGRA Collaboration],
%  ``Interferometer design of the KAGRA gravitational wave detector,''
  Phys.\ Rev.\ D {\bf 88}, no. 4, 043007 (2013)

\bibitem{TheLIGOScientific:2014jea}
J.~Aasi {\it et al.} [LIGO Scientific Collaboration],
%  ``Advanced LIGO,''
  Class.\ Quant.\ Grav.\  {\bf 32}, 074001 (2015)

  \bibitem{signal}
R. Weiss, Electromagnetically coupled broadband gravitational antenna,
LIGO Report No. LIGO-P720002, https://dcc.ligo.org/LIGO-P720002/public/main.


\bibitem{SW}
S. Weingerg, Gravitation and Cosmplogy: Principles and Applications of the General Theory of Relativity, Massachusetts Institute of Technology.

%\cite{Riles:2012yw}
\bibitem{Riles:2012yw}
  K.~Riles,
  %``Gravitational Waves: Sources, Detectors and Searches,''
  Prog.\ Part.\ Nucl.\ Phys.\  {\bf 68}, 1 (2013)
%  doi:10.1016/j.ppnp.2012.08.001
  [arXiv:1209.0667 [hep-ex]].
  %%CITATION = doi:10.1016/j.ppnp.2012.08.001;%%
  %68 citations counted in INSPIRE as of 03 Nov 2018

\bibitem{extended}
S Capozziello, M D Laurentis. Extended Theories of Gravity[J]. Physics Reports, 509 (2011) 167�21.

% \bibitem[16]{GWte}
% M. Naggiore, Gravitational Waves: Theory and Experiments, Oxford University Press (2014).

\bibitem{GAS}
H.C. Ohanian, R. Ruffini. Gravitation and Spacetime.

\bibitem{Jackson}
J. D. Jackson. Classical Electrodynamics. University of Califonia, Berkeley.

%\cite{Forward:1978zm}
\bibitem{Forward:1978zm}
  R.~L.~Forward,
  %``Wide Band Laser Interferometer Gravitational Radiation Experiment,''
  Phys.\ Rev.\ D {\bf 17}, 379 (1978).
%  doi:10.1103/PhysRevD.17.379.
  %%CITATION = doi:10.1103/PhysRevD.17.379;%%
  %103 citations counted in INSPIRE as of 03 Nov 2018
  Jean-Yves Vinet, REsearch in Astron. Astrophys. {\bf 2010} Vol. {\bf 10} No. {\bf 10}, 956.
  A.~Dirkes,
  %``Gravitational waves �A review on the theoretical foundations of gravitational radiation,''
  Int.\ J.\ Mod.\ Phys.\ A {\bf 33}, no. 14n15, 1830013 (2018)
%  doi:10.1142/S0217751X18300132
  [arXiv:1802.05958 [gr-qc]].
  %%CITATION = doi:10.1142/S0217751X18300132;%%

\bibitem{Abbott:2016blz}
  B.~P.~Abbott {\it et al.} [LIGO Scientific and Virgo Collaborations],
  %``Observation of Gravitational Waves from a Binary Black Hole Merger,''
  Phys.\ Rev.\ Lett.\  {\bf 116}, no. 6, 061102 (2016)
%  doi:10.1103/PhysRevLett.116.061102
  [arXiv:1602.03837 [gr-qc]].
  %%CITATION = doi:10.1103/PhysRevLett.116.061102;%%
  %3201 citations counted in INSPIRE as of 31 Oct 2018

%\cite{Pfenning:2000zf}
\bibitem{Pfenning:2000zf}
  M.~J.~Pfenning and E.~Poisson,
  %``Scalar, electromagnetic, and gravitational selfforces in weakly curved space-times,''
  Phys.\ Rev.\ D {\bf 65}, 084001 (2002)
%  doi:10.1103/PhysRevD.65.084001
  [gr-qc/0012057].
  %%CITATION = doi:10.1103/PhysRevD.65.084001;%%
  %53 citations counted in INSPIRE as of 03 Nov 2018
  C.~G.~Tsagas,
  %``Electromagnetic fields in curved spacetimes,''
  Class.\ Quant.\ Grav.\  {\bf 22}, 393 (2005)
%  doi:10.1088/0264-9381/22/2/011
  [gr-qc/0407080].
  %%CITATION = doi:10.1088/0264-9381/22/2/011;%%
  %82 citations counted in INSPIRE as of 03 Nov 2018


\bibitem{TE}
Bernhard J. Hoenders, Reindert Graaff, Optics Communications 255 (2005) 184.
A. Ranfagni, P. Fabeni, G. P. Pazzi, and D. Mugnai, Phys. Rev. E{\bf 48}, 1453.

\bibitem{berwaves}
Franks S. Crawford. Jr. Waves, Berkeley Physics Course - Volume 3,
University of California, Berkeley.
%\cite{Cao:2008wn}

%\cite{Culetu:2016eek}
\bibitem{Culetu:2016eek} 
  H.~Culetu,
  %``On a nonlinear gravitational wave. Geodesics,''
  PTEP {\bf 2016}, no. 12, 123E02 (2016)
%  doi:10.1093/ptep/ptw181
  [arXiv:1605.04467 [gr-qc]].
  %%CITATION = doi:10.1093/ptep/ptw181;%%
  %3 citations counted in INSPIRE as of 09 Nov 2018
  R.~J.~Slagter,
  %``Nonlinear gravitational waves as dark energy in warped spacetimes,''
%  doi:10.1142/9789813226609_0245
  arXiv:1407.7505 [gr-qc].
  %%CITATION = doi:10.1142/9789813226609_0245;%%
  %1 citations counted in INSPIRE as of 09 Nov 2018
  R.~R.~Caldwell, C.~Devulder and N.~A.~Maksimova,
  %``Gravitational wave¨CGauge field oscillations,''
  Phys.\ Rev.\ D {\bf 94}, no. 6, 063005 (2016)
%  doi:10.1103/PhysRevD.94.063005
  [arXiv:1604.08939 [gr-qc]].
  %%CITATION = doi:10.1103/PhysRevD.94.063005;%%
  %16 citations counted in INSPIRE as of 11 Nov 201


\bibitem{Cao:2008wn}
  Z.~j.~Cao, H.~J.~Yo and J.~P.~Yu,
  %``A Reinvestigation of Moving Punctured Black Holes with a New Code,''
  Phys.\ Rev.\ D {\bf 78}, 124011 (2008)
%  doi:10.1103/PhysRevD.78.124011
  [arXiv:0812.0641 [gr-qc]].
  %%CITATION = doi:10.1103/PhysRevD.78.124011;%%
  L.~W.~Ji, R.~G.~Cai and Z.~Cao,
  %``Binary black hole simulation with an adaptive finite element method III: Evolving a single black hole,''
  arXiv:1805.10642 [gr-qc].
  %%CITATION = ARXIV:1805.10642;%%
  Z.~Cao, P.~Fu, L.~W.~Ji and Y.~Xia,
  %``Binary black hole simulation with an adaptive finite element method II: Application of local discontinuous Galerkin method to Einstein equations,''
  arXiv:1805.10640 [gr-qc].
  %%CITATION = ARXIV:1805.10640;%%
  %1 citations counted in INSPIRE as of 03 Nov 2018
  %23 citations counted in INSPIRE as of 03 Nov 2018
  A.~Torres-Orjuela, X.~Chen, Z.~Cao and P.~Amaro-Seoane,
  %``Detecting the Beaming Effect of Gravitational Waves,''
  arXiv:1806.09857 [astro-ph.HE].
  %%CITATION = ARXIV:1806.09857;%%

\end{thebibliography}
\end{document}